\documentclass[reprint,amssymb, amsmath, aps, superscriptaddress, showpacs, footinbib, prb]{revtex4-1}
\usepackage{graphicx,epstopdf} 
\usepackage{amsmath}

\newcommand{\be}{\begin{equation}}
\newcommand{\ee}{\end{equation}}
\newcommand{\bea}{\begin{eqnarray}}
\newcommand{\eea}{\end{eqnarray}}
\newcommand{\bse}{\begin{subequations}}
\newcommand{\ese}{\end{subequations}}

\begin{document}

\title{Singlet ground state in the spin-$1/2$ weakly coupled dimer compound NH$_4$[(V$_2$O$_3$)$_2$(4,4$^\prime$-$bpy$)$_2$(H$_2$PO$_4$)(PO$_4$)$_2$]$\cdot$0.5H$_2$O}

\author{U. Arjun}
\affiliation{School of Physics, Indian Institute of Science Education and Research Thiruvananthapuram-695016, India}
\author{Vinod Kumar}
\affiliation{Department of Physics, Indian Institute of Technology Bombay, Mumbai-400076, India}
\author{P. K. Anjana}
\affiliation{School of Chemistry, Indian Institute of Science Education and Research Thiruvananthapuram-695016, India}
\author{A. Thirumurugan}
\affiliation{School of Chemistry, Indian Institute of Science Education and Research Thiruvananthapuram-695016, India}
\author{J. Sichelschmidt}
\affiliation{Max Planck Institut f\"{u}r Chemische Physik fester Stoffe, N\"{o}thnitzer Str.~40, 01187 Dresden, Germany}
\author{A. V. Mahajan}
\affiliation{Department of Physics, Indian Institute of Technology Bombay, Mumbai-400076, India}
\author{R. Nath}
\email{rnath@iisertvm.ac.in}
\affiliation{School of Physics, Indian Institute of Science Education and Research Thiruvananthapuram-695016, India}

\begin{abstract}
We present the synthesis and a detailed investigation of structural and magnetic properties of polycrystalline NH$_4$[(V$_2$O$_3$)$_2$(4,4$^\prime$-$bpy$)$_2$(H$_2$PO$_4$)(PO$_4$)$_2$]$\cdot$0.5H$_2$O by means of x-ray diffraction, magnetic susceptibility, electron spin resonance, and $^{31}$P nuclear magnetic resonance measurements. Temperature dependent magnetic susceptibility could be described well using a weakly coupled spin-$1/2$ dimer model with an excitation gap $\Delta/k_{\rm B}\simeq 26.1$~K between the singlet ground state and triplet excited states and a weak inter-dimer exchange coupling $J^\prime/k_{\rm B} \simeq 4.6$~K. A gapped chain model also describes the data well with a gap of about 20 K. The electron spin resonance intensity as a function of temperature traces the bulk susceptibility nicely. The isotropic Land$\acute{\rm e}$ $g$-factor is estimated to be about $g \simeq 1.97$, at room temperature. We are able to resolve the $^{31}$P NMR signal as coming from two inequivalent P-sites in the crystal structure. The hyperfine coupling constant between $^{31}$P nucleus and V$^{4+}$ spins is calculated to be $A_{\rm hf}(1) \simeq 2963$~Oe/$\mu_{\rm B}$ and $A_{\rm hf}(2) \simeq 1466$~Oe/$\mu_{\rm B}$ for the P(1) and P(2) sites, respectively. Our NMR shift and spin-lattice relaxation rate for both the $^{31}$P sites show an activated behaviour at low temperatures, further confirming the singlet ground state. The estimated value of the spin gap from the NMR data measured in an applied field of $H = 9.394$~T is consistent with the gap obtained from the magnetic susceptibility analysis using the dimer model. Because of a relatively small spin gap, NH$_4$[(V$_2$O$_3$)$_2$(4,4$^\prime$-$bpy$)$_2$(H$_2$PO$_4$)(PO$_4$)$_2$]$\cdot$0.5H$_2$O is a promising compound for further experimental studies under high magnetic fields.
\end{abstract}

\pacs{75.30.Et, 75.50.Ee, 75.10.Pq}
\maketitle

\section{\textbf{Introduction}}
Low-dimensional spin systems due to enhanced quantum fluctuations give rise to various exotic ground states which include quantum spin liquid, singlet state etc.\cite{Balents199} The ground state properties of spin systems largely depend on the lattice geometry or dimensionality and the spin value. For instance, singlet ground state is expected to occur for Haldane chains,\cite{Shimizu9835} even-leg ladders,\cite{Azuma3463} spin-Peierls systems,\cite{Hase3651} alternating spin chains,\cite{Johnston219,Ghoshray214401,Tsirlin144412,Ahmed01650} spin dimers,\cite{Sebastian617,Ruegg62,Shiramura1900} etc. An interesting aspect of the gapped spin systems is the onset of magnetic long range order (LRO) under an external magnetic field. When a magnetic field is applied, the gap between the singlet ($S=0$) and the degenerate triplet ($S=1$) excited states gets reduced. At a critical value of field, the gap is completely closed and one of the split triplet state intercepts the singlet ground state at which a three dimensional (3D) antiferromagnetic LRO emerges. This fascinating field induced phenomenon is known as Bose-Einstein Condensation (BEC) of triplons.\cite{Nikuni2000} Notable examples showing BEC are BaCuSi$_2$O$_6$ (Ref.~\onlinecite{Sebastian617}), Sr$_3$Cr$_2$O$_8$ (Ref.~\onlinecite{Aczel207203}), and TlCuCl$_3$ (Ref.~\onlinecite{Ruegg62}). In addition to BEC, spin gap materials also feature other non-trivial properties. For instance, SrCu$_2$(BO$_3$)$_2$ is known as an orthogonal dimer system with the Shastry-Sutherland lattice which shows Wigner crystallization of magnons and magnetization plateaus under an external magnetic field.\cite{Kageyama3168, Kodama395}
\begin{figure*}
\includegraphics[scale=0.07]{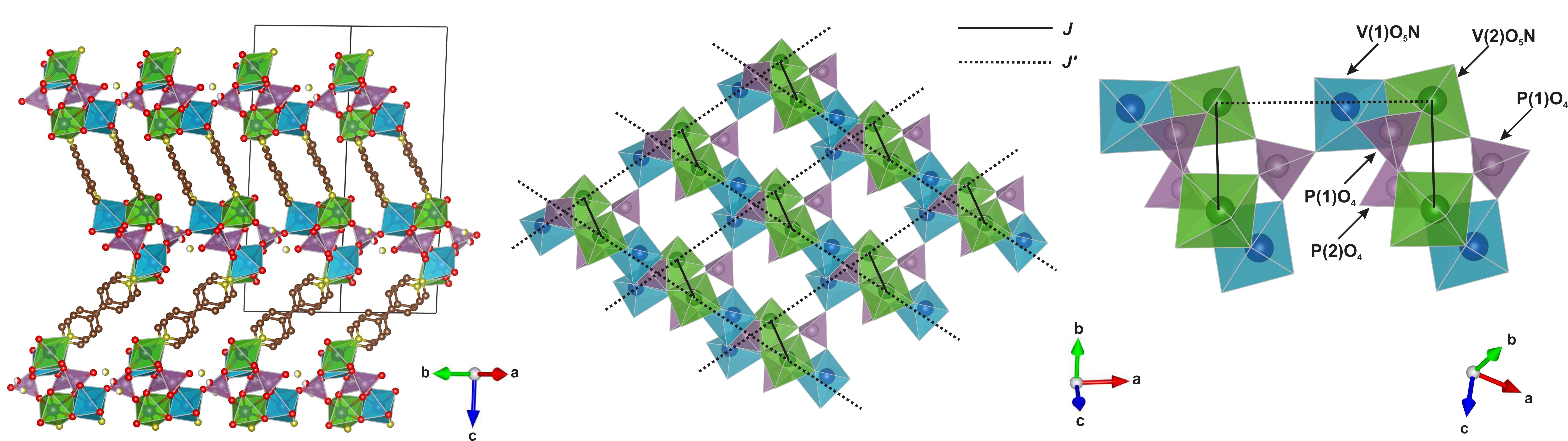}\\
\caption{Left panel: crystal structure of VP-$bpy$ viewed along [110] direction. Vanadium phosphate layers are connected by the long 4,4$^\prime$-bipyridine ligands. Middle panel: a section of the vanadium phosphate layer composed of weakly coupled dimers in the $ab$-plane is shown. The solid line represents the dimer units while the dotted line represents the direction of alternating cross chains. Right panel: two dimer units with intra-dimer coupling ($J$) via V(2)-P-O-P-V(2) path and inter-dimer coupling ($J^\prime$) via V(2)-O-V(1)-P-O-V(2) path are shown by solid and dotted lines, respectively.}
\label{Fig1}
\end{figure*}

Efficient experimental studies require appropriate compounds with weak exchange couplings that allow exploring the full temperature versus field phase diagram and the associated quantum phase transitions. Therefore, the primary goal is to search for new spin gap systems displaying field induced effects under the accessible magnetic fields in the laboratory. In this respect, organic compounds with spin-$1/2$ metal ions are crucial since they have a low energy scale of the exchange couplings compared to inorganic compounds. A few spin-$1/2$ organic compounds are reported to show a singlet ground state.\cite{Bonner248} Some of the examples are Cu(NMI)$_2$Br$_2$ [Ref.~\onlinecite{Smit229}], Cu(4MP)$_2$Cl$_2$, Cu-HTS [Ref.~\onlinecite{Hatfield1985}], Cu-OTS [Ref.~\onlinecite{Hatfield1985}], Cu(NO$_3$)$_2$$\cdot$2.5H$_2$O [Ref.~\onlinecite{Bonner248}], and VODPO$_4$$\cdot\frac{1}{2}$D$_2$O [Ref.~\onlinecite{Tennant4998}]. All of them except VODPO$_4$$\cdot\frac{1}{2}$D$_2$O exhibit alternating chain behaviour whereas only in VODPO$_4$$\cdot\frac{1}{2}$D$_2$O, the weakly coupled dimers are responsible for the singlet ground state.

In this paper, we present the structural and magnetic properties of the spin-$1/2$ dimer compound NH$_4$[(V$_2$O$_3$)$_2$(4,4$^\prime$-$bpy$)$_2$(H$_2$PO$_4$)(PO$_4$)$_2$]$\cdot$0.5H$_2$O (abbreviated as VP-$bpy$) investigated via x-ray diffraction (XRD), magnetic susceptibility, electron spin resonance (ESR), and $^{31}$P nuclear magnetic resonance (NMR) measurements. It crystallizes in a monoclinic space group $C2/c$ (no.~15) with lattice parameters $a = 12.6354(8)$~\AA, $b = 9.9786(6)$~\AA, $c = 23.369(1)$~\AA, $\alpha = 90^{\circ}$, $\beta = 92.713(1)^{\circ}$, and $\gamma = 90^{\circ}$.\cite{Hung3929} The crystal structure of VP-$bpy$, as shown in Fig.~\ref{Fig1}, contains two inequivalent vanadium [V(1) and V(2)] and two inequivalent phosphorous [P(1) and P(2)] atoms. The number of P(1) atoms is double the number of P(2) atoms present in the unit cell. The oxidation states of V(1) and V(2) are 5+ (non-magnetic) and 4+ (magnetic, spin-$1/2$), respectively. Using the structural data reported in Ref.~\onlinecite{Hung3929}, bond valence sum calculations\cite{Brese192,*Han11116,*Brown6858} support the assignment of 4+ and 5+ oxidation states to V(2) and V(1) atoms, respectively. Each vanadium atom is bonded to one N and five O atoms to form distorted VO$_5$N octahedra. Magnetic dimers are formed by the corner sharing of V(2) octahedra via two P(1)O$_4$ and one P(2)O$_4$ tetrahedra. In order to understand how strongly P atoms are coupled to V$^{4+}$ spins, the atomic distances and the corresponding angles within a dimer unit are tabulated in Table~\ref{TableI}.
\begin{table}
	\noindent
	\caption{Atomic distances and the corresponding angles for the interaction paths V(2)-O(2)-P(1)-O(4)-V(2) and V(2)-O(5)-P(2)-O(5)-V(2), within one dimer unit.}
	\label{TableI}
	\begin{tabular}{c|c|c}
		\hline \hline
		& Bond length (\AA) & Angle (degree) \\\hline
		P(1) & V(2)-O(2) = 1.958 & $\angle$ V(2)-O(2)-P(1) = 144.53  \\
		& O(2)-P(1) = 1.516 & $\angle$ P(1)-O(4)-V(2) = 130.71 \\
		& P(1)-O(4) = 1.535 & \\
		& O(4)-V(2) = 1.991 & \\\hline
		P(2) & V(2)-O(5) = 2.29 & $\angle$ V(2)-O(5)-P(2) = 135.32 \\
		& O(5)-P(2) = 1.507 & $\angle$ P(2)-O(5)-V(2) = 135.32 \\
		& P(2)-O(5) = 1.507 & \\
		& O(5)-V(2) = 2.29 & \\
		\hline \hline
	\end{tabular}
\end{table}
From Table~\ref{TableI} it is clear that P(2) is located symmetrically while P(1) is located asymmetrically between the V(2) atoms. But the distance between V(2) and P(1) atoms via O(2) or O(4) is smaller than the distance between V(2) and P(2) atoms via O(5). Therefore, it is expected that the P(1) atom is more strongly coupled to the V(2)$^{4+}$ ions than the P(2) atom. The dimers are connected to each other through a long V(2)-O-V(1)-O-P-O-V(2) pathway resulting in a two-dimensional (2D) vanadium phosphate layer. Alternatively, the weakly interacting array of spin dimers can also be viewed as alternating cross chains (middle panel of Fig.~\ref{Fig1}) running nearly perpendicular to each other. The unit cell contains two such vanadium phosphate layers and they are well separated ($\sim 8$~\AA) from each other. Subsequently, the layers are further linked through long 4,4$^\prime$-bipyridine pillars of length $\sim 14$~\AA~to generate a three-dimensional (3D) framework.

Our magnetic measurements suggest a weakly coupled spin-$1/2$ dimer behaviour of the compound with a non-magnetic or singlet ground state. Equivalently, the gapped Heisenberg chain model fits the data equally well. The NMR shift and spin-lattice relaxation rate also show an exponential decrease at low temperatures, further confirming the opening of a spin gap.

\section{\textbf{Experimental details}}
Single crystals of VP-$bpy$ were synthesized following the hydrothermal reaction route. The reaction was carried out in a Teflon-lined stainless steel bomb with an internal volume of 20~mL. The reaction of V$_2$O$_5$ (0.3~mmol), 4,4$^\prime$-bipyridine (0.9~mmol), H$_3$PO$_4$ (0.104~mL), NH$_4$OH(aqueous) (0.1~mL), and H$_2$O (8~mL) at 160~$^{\circ}$C for 3 days produced small dark green crystals. Our single crystal XRD (Bruker APEX-II machine with MoK$_{\rm \alpha1}$ radiation, $\lambda_{\rm avg} \simeq 0.7107$~\AA) experiments on a good quality crystal were performed at two different temperatures $T=300$~K and 150~K. Both the data sets confirm the monoclinic structure with space group $C2/c$ and the obtained lattice parameters are as follows: [$a=12.6332(5)$~\AA, $b=9.9336(4)$~\AA, $c=23.3407(8)$~\AA,
$\alpha=90^{\circ}$, $\beta = 92.7060(10)^{\circ}$, $\gamma=90^{\circ}$] and [$a=12.6300(5)$~\AA, $b=9.8928(4)$~\AA, $c=23.2371(8)$~\AA, $\alpha=90^{\circ}$, $\beta = 92.8150(10)^{\circ}$, $\gamma=90^{\circ}$], respectively.

\begin{figure}[h]
\includegraphics{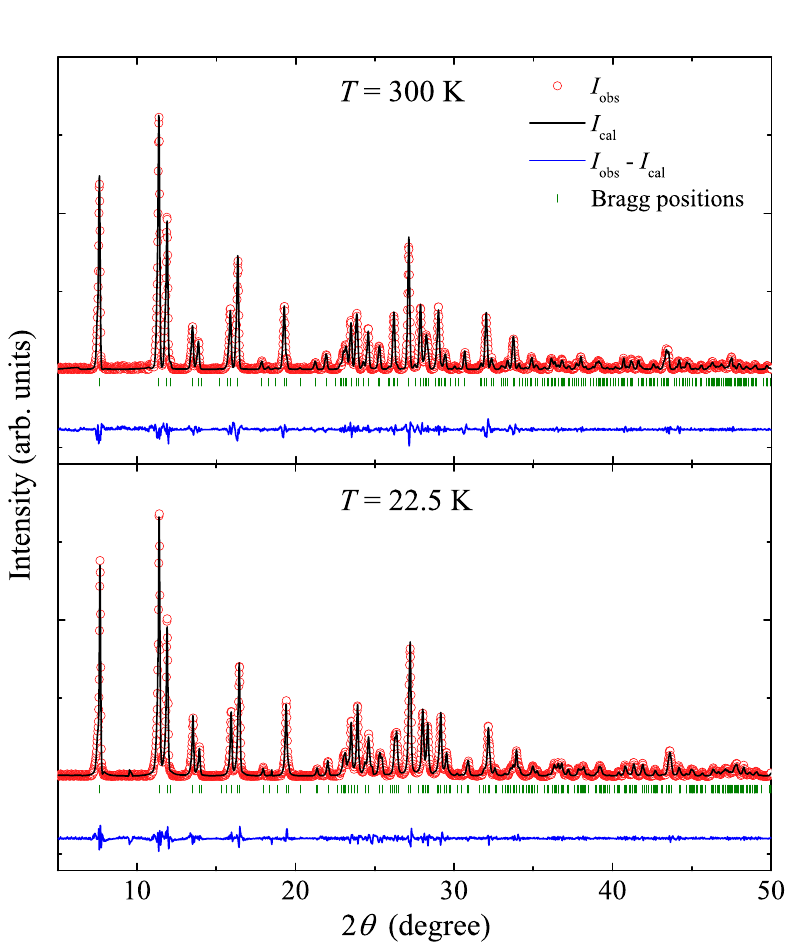}\\
\caption{Powder x-ray diffraction patterns (open circles) at $T=300$~K (top panel) and $22.5$~K (bottom panel) for VP-$bpy$. The solid red line represents the calculated pattern using Le-Bail fit, with the vertical bars showing the expected Bragg peak positions, and the lower solid blue line representing the difference between the observed and calculated intensities.}
\label{Fig2}
\end{figure}
In order to further confirm the phase purity, powder XRD was performed on the crushed powder sample at different temperatures using a PANalytical (Cu $K_{\alpha}$ radiation, $\lambda_{\rm avg} \simeq 1.5418$~\AA) powder diffractometer. Le-Bail fit of the observed powder XRD patterns was performed using the \verb FullProf  package.\cite{Carvajal55} The initial structural parameters for this purpose were taken from Ref.~\onlinecite{Hung3929}. Figure~\ref{Fig2} shows the powder XRD data of VP-$bpy$ at room temperature ($T = 300$~K) and at $T = 22.5$~K along with the calculated patterns. All the peaks could be fitted using the monoclinic ($C2/c$) structure. The obtained best fit parameters are [$a = 12.6004(2)$~\AA, $b = 9.9346(2)$~\AA, $c = 23.3378(5)$~\AA, $\beta = 92.673(2)^{\circ}$, $V_{\rm cell} \simeq 2918.26$~\AA$^3$, and the goodness-of-fit $\chi^2 \simeq 15$] and [$a=12.6150(6)$~\AA,
$b=9.8764(4)$~\AA, $c=23.162(1)$~\AA, $\beta=92.909(3)^{\circ}$, $V_{\rm cell} \simeq 2882.1$~\AA$^3$, and $\chi^2 \simeq 14$] for $T = 300$~K and $22.5$~K, respectively. The obtained lattice parameters at room temperature are in reasonable agreement with the single crystal XRD results and also with the previous report.\cite{Hung3929}
No structural transition was observed down to $22.5$~K but the unit cell volume was found to be slightly smaller at low temperature.

Magnetization $M$ was measured as a function of temperature (2.1~K~$\leq T \leq$~380~K) using the vibrating sample magnetometer (VSM) attachment to the Physical Property Measurement System [PPMS, Quantum Design]. The ESR experiments were carried out on a fine-powdered sample with a standard continuous-wave spectrometer between 3\,K and 300\,K. We measured the power $P$ absorbed by the sample from a transverse magnetic microwave field (X-band, $\nu\simeq 9.4$\,GHz) as a function of the external magnetic field $H$. A lock-in technique was used to improve the signal-to-noise ratio which yields the derivative of the resonance signal $dP/dB$.

The NMR experiments on $^{31}$P nucleus (nuclear spin $I = 1/2$ and gyromagnetic ratio $\gamma/2\pi=17.235$~MHz/T) were carried out using pulsed NMR techniques in a fixed magnetic field of 9.394~T. The $^{31}$P NMR spectra as a function of temperature were obtained by doing Fourier transform of the echo signal. In the temperature range $10-50$~K where the $^{31}$P line is broad, spectra were obtained by plotting the echo integral as a function of frequency, at each temperature. The NMR shift $K(T)=[\nu(T)-\nu_{\rm ref}]/\nu_{\rm ref}$ was determined by measuring the resonance frequency $\nu(T)$ of the sample with respect to the standard H$_3$PO$_4$ sample (resonance frequency $\nu_{\rm ref}$). The $^{31}$P nuclear spin-lattice relaxation rate ($1/T_1$) was measured using a conventional saturation pulse sequence.

\section{\textbf{Results and discussion}}
\subsection{\textbf{Magnetic susceptibility}}
\begin{figure}[h]
\includegraphics{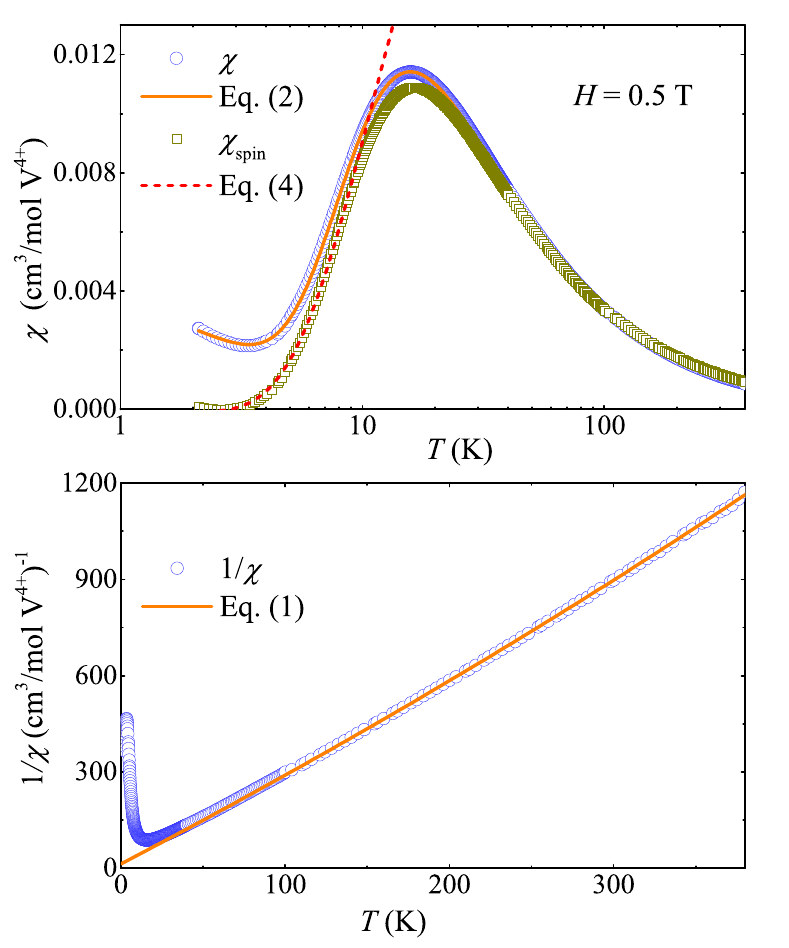}\\
\caption{Upper panel: $\chi(T)$ measured at $H = 0.5$~T. The solid and dotted lines represent the fit of $\chi(T)$ and $\chi_{\rm spin}(T)$ using Eqs.~\eqref{chi_alt} and \eqref{gap}, respectively. Lower panel: $1/\chi$ vs $T$ and the solid line is the CW fit using Eq.~\eqref{cw}.}
\label{Fig3}
\end{figure}

Temperature dependent magnetic susceptibility $\chi(T)$~$(\equiv M/H)$ measured in an applied field of $H = 0.5$~T is shown in the upper panel of Fig.~\ref{Fig3}. With decreasing temperature, $\chi(T)$ increases in a Curie-Weiss manner as is expected in the paramagnetic regime and then shows a broad maximum at $T_\chi^ {\rm {max}} \simeq 15$~K, indicative of short range magnetic order which is also a hallmark of low dimensionality. Below $T^{\rm max}_{\rm \chi}$, $\chi(T)$ decreases rapidly which is a signature of the opening of a spin gap. At very low temperatures, it shows an upturn, likely due to extrinsic paramagnetic impurities and/or defects present in the sample. There is no clear indication of any magnetic LRO down to 2.1~K.

To extract the magnetic parameters, $\chi(T)$ at high temperatures was fitted by the following expression
\begin{equation}\label{cw}
\chi(T) = \chi_0 + \frac{C}{T - \theta_{\rm CW}},
\end{equation}
where $\chi_0$ is the temperature independent contribution consisting of core diamagnetic susceptibility ($\chi_{\rm core}$) of the core electron shells and Van-Vleck paramagnetic susceptibility ($\chi_{\rm VV}$) of the open shells of the V$^{4+}$ ions present in the sample. The second term in Eq.~\eqref{cw} is the Curie-Weiss (CW) law with the CW temperature $\theta_{\rm CW}$ and Curie constant $C = N_{\rm A} \mu_{\rm eff}^2/3k_{\rm B}$. Here $N_{\rm A}$ is the Avogadro number, $k_{\rm B}$ is the Boltzmann constant, $\mu_{\rm eff} = g\sqrt{S(S+1)}$$\mu_{\rm B}$ is the effective magnetic moment, $g$ is the Land$\acute{\rm e}$ $g$-factor, $\mu_{\rm B}$ is the Bohr magneton, and $S$ is the spin quantum number.
Our fit in the temperature range 150~K to 380~K (lower panel of Fig.~\ref{Fig3}) yields $\chi_0 \simeq -1.11 \times 10^{-4}$~cm$^3$/mol-V$^{4+}$, $C \simeq 0.373$~cm$^3$K/mol-V$^{4+}$, and $\theta_{\rm CW} \simeq -4.8$~K. From the value of $C$, the effective moment was calculated to be $\mu_{\rm eff} \simeq 1.72 \mu_{\rm B}$/V$^{4+}$ which is in close agreement with the expected spin-only value of 1.73~$\mu_{\rm B}$ for $S = 1/2$, assuming $g = 2$. The negative value of $\theta_{\rm CW}$ indicates that the dominant exchange couplings between V$^{4+}$ ions are antiferromagnetic in nature.

In order to understand the spin-lattice and to estimate the exchange couplings, $\chi(T)$ was fitted by the following expression
\begin{equation}\label{chi_alt}
\chi(T) = \chi_0 + \frac {C_{\rm imp}}{T-\theta_{\rm imp}}+\chi_{\rm spin}.
\end{equation}
Here, $C_{\rm imp}$ represents the Curie constant corresponding to the impurity spins and $\theta_{\rm imp}$ provides an effective interaction between impurity spins. $\chi_{\rm spin}$ is the expression for spin susceptibility of interacting spin-$1/2$ dimers which has the form\cite{Johnston1,Singh012407}
\begin{equation}\label{dimer}
	\chi_{\rm spin} = \frac {N_{\rm A} g^2 \mu_{\rm B}^2}{k_{\rm B} T(3+e^{{\Delta}/k_{\rm B}T}+J^\prime/k_{\rm B}T)}.
\end{equation}
Here, $\Delta/k_{\rm B}$ represents the spin gap between the singlet ground state and the spin-1 triplet excited states which is equal to the intra-dimer exchange coupling ($J/k_{\rm B}$) and $J^\prime/k_{\rm B}$ is the average inter-dimer exchange coupling. The expression in Eq.~\eqref{dimer} is not exact because a mean-field type approximation is used in order to introduce $J^\prime/k_{\rm B}$ in the expression. The second term (CW term) is included in Eq.~\eqref{chi_alt} to account for the low temperature upturn in $\chi(T)$. As shown in the upper panel of Fig.~\ref{Fig3}, Eq.~\eqref{chi_alt} fits very well to the $\chi(T)$ data over the whole temperature range. While fitting, the value of $g$ was fixed to $g \simeq 1.97$, obtained from the ESR experiments (discussed later). The obtained best fit parameters are $\chi_0\simeq-1.129 \times 10^{-4}$~cm$^3$/mol-V$^{4+}$, $C_{\rm imp}\simeq0.011$~cm$^3$K/mol-V$^{4+}$, $\Delta/k_{\rm B} \simeq 26.1$~K, $J^\prime/k_{\rm B} \simeq 4.6$~K, and $\theta_{\rm imp} \simeq -2$~K. This value of $C_{\rm imp}$ corresponds to a spin concentration of nearly 2.7~\%, assuming the impurity spins to be $S = 1/2$. The critical field for closing a gap of $\Delta/k_{\rm B} \simeq 26.1$~K is estimated to be $H_{\rm c} = \frac{\Delta}{g \mu_{\rm B}} \simeq 19.6$~T.


The agreement of Eq.~\eqref{chi_alt} with our experimental $\chi(T)$ data over the whole temperature range suggests weakly coupled dimer behaviour of VP-$bpy$ which is consistent with the predictions based on the structural data.
To further demonstrate the gapped nature, $\chi_0 + \frac {C_{\rm imp}}{T-\theta_{\rm imp}}$ was subtracted from the experimental $\chi(T)$ data and the obtained $\chi_{\rm spin}$ is plotted in the same figure (Fig.~\ref{Fig3}). Below the broad maximum, $\chi_{\rm spin}$ clearly shows an exponential decrease towards zero, corroborating the singlet ground state. 
We found that the low temperature ($T < \Delta/k_{\rm B}$) $\chi(T)$ data could be fitted well by the following expression valid for a gapped Heisenberg 1D chain,\cite{Sachdev943,Damle8307}
\begin{equation}
\label{gap}
\chi = A\sqrt{\frac{\Delta}{k_{\rm B}T}}e^{-\Delta/k_{\rm B}T},
\end{equation}
where $A$ is a constant. Our fit to the $\chi_{\rm spin}(T)$ data below 12~K yields $\Delta/k_{\rm B} \simeq 20$~K which is slightly smaller than the value obtained from the interacting dimer model fit. The critical field corresponding to the gap closing would be $H_{\rm c} \simeq 15$~T.




\subsection{ESR}
\begin{figure}[h]
	\includegraphics{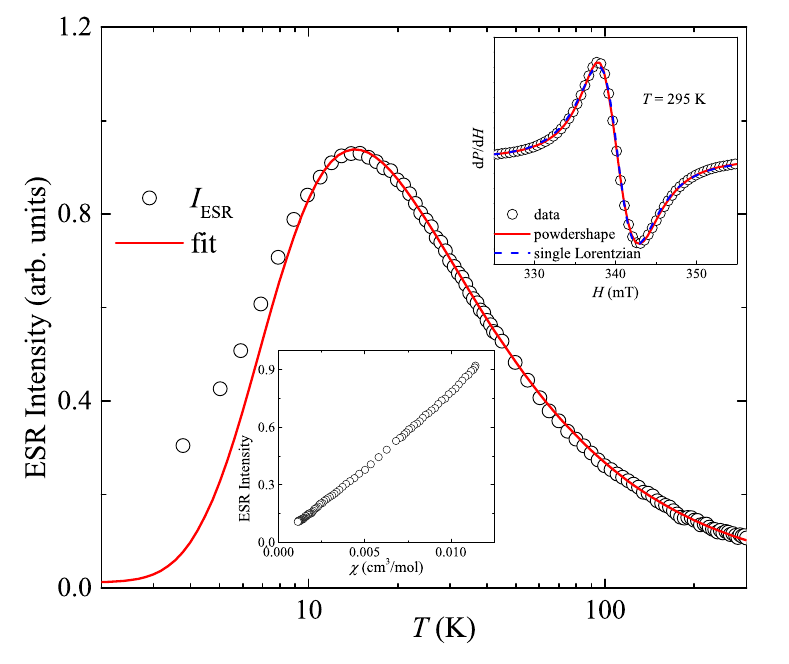}\\
	\caption{Temperature dependent ESR intensity, $I_{\rm ESR}(T)$, obtained by integration of the ESR spectra of the polycrystalline VP-$bpy$ sample. The solid line represents the fit described in the text. The upper inset shows a typical spectrum (symbols) together with the fits using a single Lorentzian function (dashed line) and a powder averaged Lorentzian shape (solid line) for a uniaxial $g$-factor anisotropy. The bottom inset shows the $I_{\rm ESR}$ vs $\chi$ in the temperature range, 15~K to 295~K.}
	\label{Fig4}
\end{figure}
Results of the ESR experiment on the VP-$bpy$ powder sample are presented in Fig.~\ref{Fig4}. In the upper inset of Fig.~\ref{Fig4}, a typical ESR spectrum at room temperature is shown. We tried to fit the spectra using a single Lorentzian function and a powder-averaged Lorentzian line shape. Both fits reproduce the spectral shape very well at $T=295$\,K yielding an average $g$-factor of $\sim 1.973$ and anisotropic (parallel $g_{\parallel} \simeq 1.982$ and perpendicular $g_{\perp} \simeq 1.967$) components, respectively. For the latter case, the isotropic
$g$ $\left[ =\sqrt{(g^{2}_{\parallel}+2g^{2}_{\perp})/3}\right]$ value was calculated to be $\sim 1.972$ which is nearly same as the value obtained from the former fit. A reduced value of $g$, typical for V$^{4+}$, has been found from the ESR experiments on AgVOAsO$_4$ (Ref.~\onlinecite{Tsirlin144412}), Sr$_2$V$_3$O$_9$ (Ref.~\onlinecite{Ivanshin064404}), Pb$_2$VO(PO$_4$)$_2$ (Ref.~\onlinecite{Forster180401}), SrZnVO(PO$_4$)$_2$ (Ref.~\onlinecite{Forster014032}), and Zn$_2$VO(PO$_4$)$_2$ (Ref.~\onlinecite{Yogi024413}). The value of $g$ is found to be temperature independent (not shown).

The ESR intensity $I_{\rm ESR}(T)$ obtained by integrating the whole spectrum as a function of temperature is plotted in Fig.~\ref{Fig4}. It is found to increase with decreasing temperature and then shows a broad maximum at around 15~K, below which it decreases rapidly. The overall shape of the curve resembles the bulk $\chi(T)$ data (Fig.~\ref{Fig3}) and the rapid decrease at low temperatures is a clear signature of the opening of a spin gap. In an attempt to see how $I_{\rm ESR}$ scales with $\chi$, we plotted $I_{\rm ESR}$ vs $\chi$ with temperature as an implicit parameter (see, lower inset of Fig.~\ref{Fig4}). A nearly linear behavior down to 15\,K reflects that $I_{\rm ESR}(T)$ tracks $\chi(T)$ of the V$^{4+}$ spins probed by ESR. Thus, one can get an estimation of the exchange couplings from the fitting of $I_{\rm ESR}(T)$ data by a coupled dimer model. We fitted the data by $I_{\rm ESR}(T) = A+B\times \chi_{\rm spin}$, where $A$ and $B$ are arbitrary constants and $\chi_{\rm spin}$ is the expression of spin susceptibility for coupled dimer model as given in Eq.~\eqref{dimer}. Our fitting in the temperature range 8~K to 300~K fixing $g \simeq 1.97$ (obtained from ESR spectra) and $J^\prime/k_{\rm B} \simeq 4.6$~K [from $\chi(T)$] yields $\Delta/k_{\rm B} \simeq 23$\,K which is close to the value estimated from the $\chi(T)$ analysis. The deviation of experimental data from the fit below 8~K is likely due to the fact that the ESR intensity also probes the magnetization of the paramagnetic impurity spins at low temperatures similar to the bulk $\chi(T)$ data.


\subsection{\textbf{$^{31}$P NMR}}
\subsubsection{\textbf{NMR shift}}
\begin{figure}[h]
\includegraphics{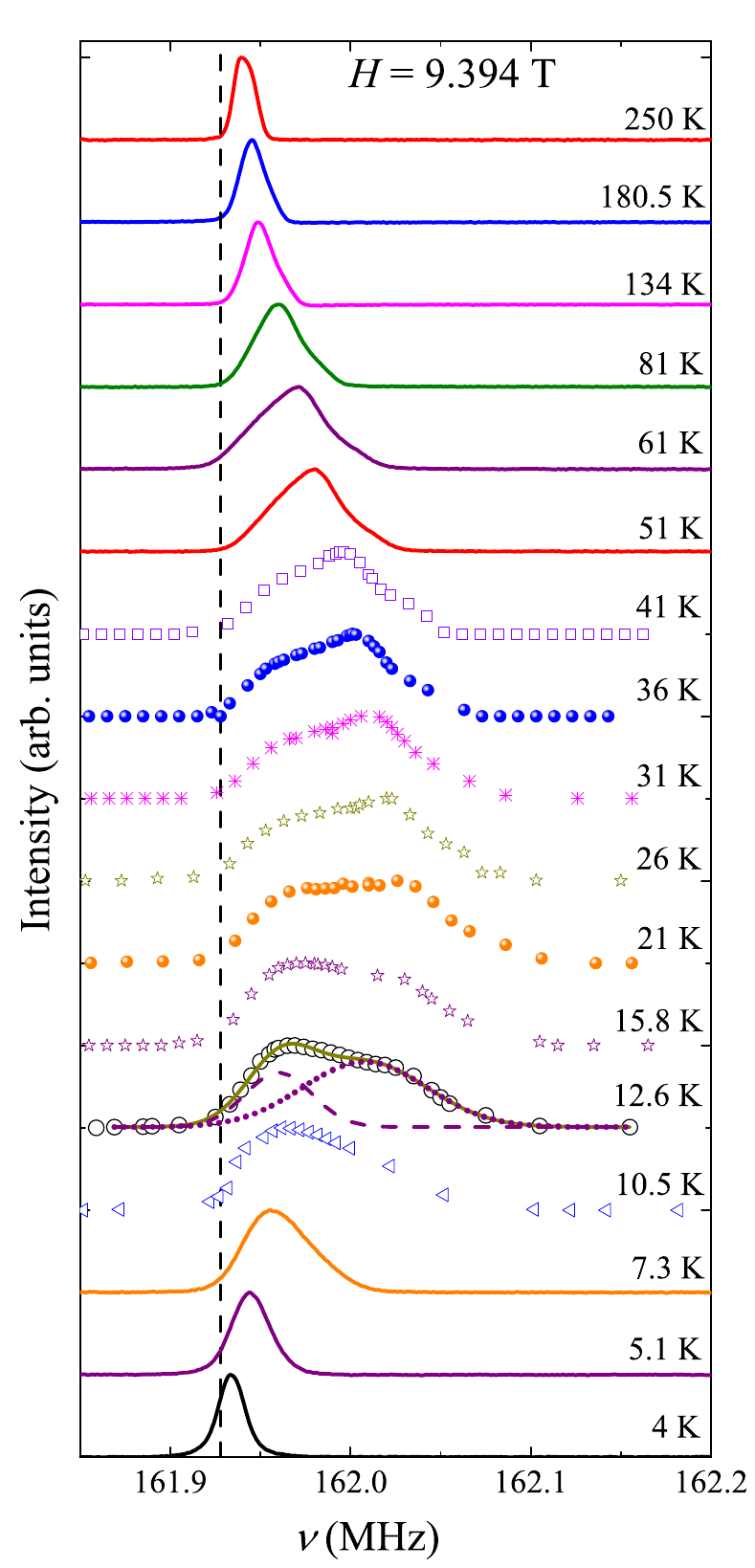}\\
\caption{Temperature dependent frequency sweep $^{31}$P NMR spectra of VP-$bpy$ measured at $H = 9.394$~T.
The double Gaussian fit to the spectrum is shown at $T = 12.6$~K. The dotted and dashed lines correspond to the P(1) and P(2) sites, respectively. The solid line represents the total fit. The ratio of areas under P(1) and P(2) spectra is nearly 2:1. The vertical dashed line corresponds to the $^{31}$P resonance position of the reference H$_3$PO$_4$ sample.}
\label{Fig5}
\end{figure}
NMR is a powerful local probe to study the static and dynamic properties of a spin system. In VP-$bpy$, the $^{31}$P nuclei are inductively coupled to the magnetic spins. Therefore, the low-lying excitations of the V$^{4+}$ spins can be probed by measuring $^{31}$P NMR spectra, NMR shift, and spin-lattice relaxation time.
Since $^{31}$P has nuclear spin $I=1/2$, one would expect a single spectral line from $^{31}$P NMR corresponding to one allowed transition. Figure~\ref{Fig5} displays representative spectra at some selected temperatures. We indeed observed a single and narrow spectral line at high temperatures but the line shape is asymmetric. The asymmetric line shape reflects either anisotropy in spin susceptibility and/or anisotropy of the hyperfine coupling.
As the temperature is lowered, the line broadens and exhibits a shoulder for $T\lesssim 50$~K. Such a line shape clearly reflects that two inequivalent $^{31}$P sites feel the hyperfine field of V$^{4+}$ spins differently, similar to that reported for (VO)$_2$P$_2$O$_7$.\cite{Kikuchi6731} With further decrease in temperature ($T \lesssim 12$~K), the line width decreases significantly and it becomes a single narrow line at very low temperatures.

\begin{figure}[h]
\includegraphics{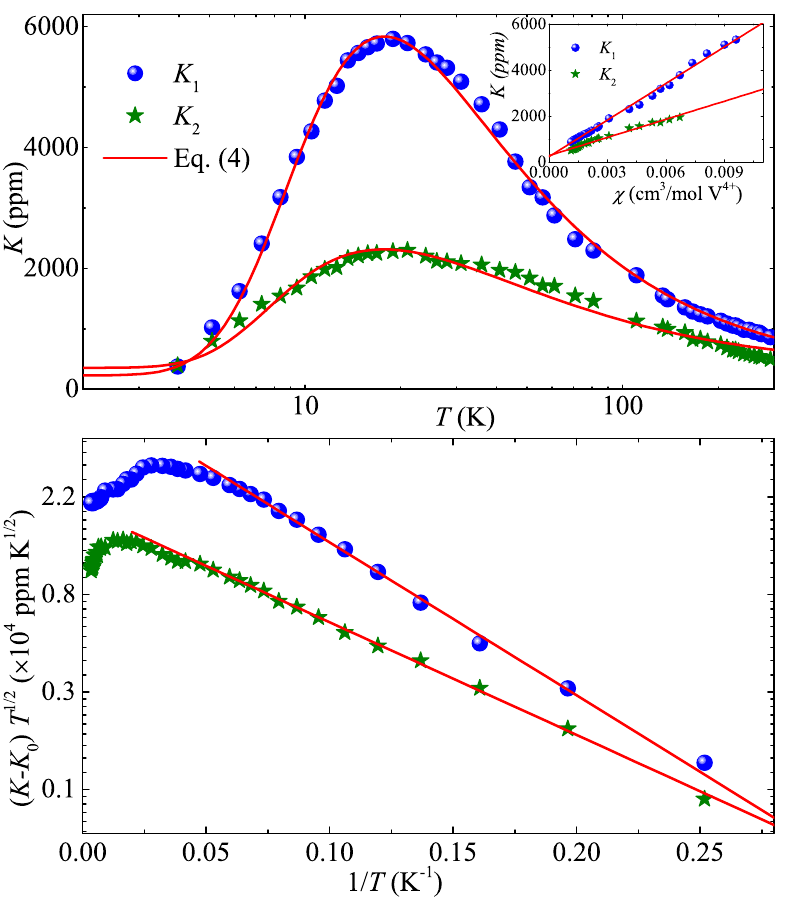}\\
\caption{Upper panel: temperature dependent $^{31}$P NMR shift ($K_1$ and $K_2$) for both the $^{31}$P sites as a function of temperature. The solid lines are the fits of $K(T)$ by Eq.~\eqref{K_chi}. Inset: $K$ vs $\chi$ (measured at 9~T) with temperature as an implicit parameter. The solid lines are the linear fits. Lower panel: $(K-K_0)T^{1/2}$ vs $1/T$ for both the P sites. The solid lines indicate the activation law $(K-K_0)T^{1/2} \propto e^{-\Delta/k_{\rm B}T}$.}
\label{Fig6}
\end{figure}
As demonstrated in Fig.~\ref{Fig5}, both the shoulder positions are found to shift with temperature. The temperature-dependent NMR shift $K(T)$ for both the $^{31}$P sites was extracted by fitting the spectra to a sum of two Gaussian functions. Since the number of P(1) atoms is double the number of P(2) atoms present in the unit cell, one expects the area under the spectrum corresponding to the P(1) site to be double to that of the P(2) site. Thus, the right hand side shoulder with the larger area corresponds to the P(1) site whereas the left hand side shoulder with the smaller area corresponds to the P(2) site. While fitting the spectra at different temperatures, we kept the constraint that the area corresponding to P(1) is twice that of P(2). Figure~\ref{Fig6} displays the extracted $K$s. Similar to $\chi(T)$, $K(T)$ for both the $^{31}$P sites passes through a broad maximum at around 18~K, which indicates low-dimensional short-range ordering. The absolute value of $K_1(T)$ [for P(1) site] is larger and strongly temperature dependent compared to $K_2(T)$ [for P(2) site] due to the difference in hyperfine couplings. This is of course consistent with our predictions based on the crystal structure where P(1) is expected to be coupled strongly to the V$^{4+}$ spins compared to the P(2) nuclei.
At low temperatures, $K(T)$ for both the P sites decreases rapidly towards zero. This sharp fall in $K(T)$ clearly signifies the reduction of spin susceptibility of V$^{\rm 4+}$ spins and opening of a spin gap between the singlet ($S=0$) ground state and triplet ($S=1$) excited states.

As $K(T)$ is a direct measure of the spin susceptibility $\chi_{\rm spin}(T)$, one can write
\begin{equation}\label{K_chi}
K(T) = K_0 + \frac{A_{\rm hf}}{N_{A}}\chi_{\rm spin},
\end{equation}
where $K_0$ is the temperature-independent chemical shift and $A_{\rm hf}$ is the total hyperfine coupling between $^{31}$P nuclei and V$^{4+}$ spins. $A_{\rm hf}$ includes contributions from transferred hyperfine coupling and the nuclear dipolar coupling, both of which are temperature independent.
From Eq.~\eqref{K_chi}, $A_{\rm hf}$ can be calculated from the slope of the linear $K$ vs $\chi$ plot with temperature as an implicit parameter. Inset of Fig.~\ref{Fig6} presents the $K$ vs $\chi$ plots for both the $^{31}$P sites. The $\chi(T)$ data used in Fig.~\ref{Fig6} were measured at 9~T, which is close to the field at which our NMR experiments were performed. Clearly, our $K$ vs $\chi$ plot for the P(1) site is a straight line down to 27~K. The data for $T \geq 27$~K were fitted well to a linear function and the slope of the fit yields $A_{\rm hf}(1) \simeq 2963$~Oe/$\mu_{\rm B}$. On the other hand for the P(2) site, a linear fit of the $K$ vs $\chi$ plot down to 45~K gives $A_{\rm hf}(2) \simeq 1466$~Oe/$\mu_{\rm B}$.
These values of $A_{\rm hf}$ are of the same order of magnitude as those reported for $^{31}$P NMR in uniform one-dimensional (1D) spin chains (Ba,Sr)$_2$Cu(PO$_4$)$_2$ and K$_2$CuP$_2$O$_7$.\cite{Nath174436,Nath134451}


It is to be noted that the NMR shift directly probes $\chi_{\rm spin}$ and is free from impurity contributions. Therefore, in low dimensional spin systems, $K(T)$ data are often used for reliable estimation of magnetic parameters instead of the bulk $\chi(T)$.
For a tentative estimation, we fitted the $K(T)$ data using Eq.~\eqref{K_chi} over the whole temperature range, taking $\chi_{\rm spin}$ for the interacting spin-$1/2$ dimer model [Eq.~\eqref{dimer}]. To minimize the number of fitting parameters, $A_{\rm hf}$ and $g$ were fixed to the values obtained from the $K$ vs $\chi$ analysis and ESR experiments, respectively. The resultant fitting parameters are ($K_0 \simeq 228$~ppm, $\Delta/k_{\rm B} \simeq 28.4$~K, and $J^\prime/k_{\rm B} \simeq -3.7$~K) and ($K_0 \simeq 353$~ppm, $\Delta/k_{\rm B} \simeq 28.3$~K, and $J^\prime/k_{\rm B} \simeq 54$~K) for the P(1) and P(2) sites, respectively. Though the obtained values of $\Delta/k_{\rm B}$ are close to those obtained from the low-field $\chi(T)$ analysis, the $J^\prime/k_{\rm B}$ values appear to be unphysical. Especially for the shift $K_2(T)$ [for the P(2) site], there is a large uncertainty due to the large width and small absolute variation. Further note that the expression in Eq.~\eqref{dimer} is applicable for the estimation of $\Delta/k_{\rm B}$ and $J^\prime/k_{\rm B}$ from the zero field $\chi(T)$ data only. The gap is expected to decrease to about 14~K in 9.394~T so the obtained value of $\Delta/k_{\rm B}$ appears incongruous.


Once again, we could fit the low-temperature shift data to $K(T) = K_0+A\sqrt{\frac{\Delta}{k_{\rm B}T}}e^{-\Delta/k_{\rm B}T}$. Such an expression has been used to describe the susceptibility of gapped 1D systems, however in the low-field limit.\cite{Sachdev943,Damle8307}
The obtained values are ($K_{0} \simeq -347$~ppm, $A \simeq 16738$~ppm, and $\Delta/k_{\rm B} \simeq 15.6$~K) and ($K_{0} \simeq -110.8$~ppm, $b \simeq 5687.7$~ppm, and $\Delta/k_{\rm B} \simeq 12$~K) for P(1) and P(2) sites, respectively. The results of the fit are shown in the lower panel of Fig.~\ref{Fig6} where we have plotted $(K-K_{0})T^{1/2}$ vs $1/T$. The $y$-axis is shown in log scale in order to highlight the linear behaviour in the gapped region. These values of  $\Delta/k_{\rm B}$ are higher than what is expected ($\sim 7.5$~K for the gapped Heisenberg chain) at $H = 9.394$~T, assuming a linear decrease of the gap with field from the zero-field value.\footnote{Note that the spin gap estimated from the $\chi(T)$ analysis (using dimer model) is $\Delta/k_{\rm B} \simeq 26.1$~K, at zero field and the corresponding critical field of gap closing would be $H_{\rm c} \simeq 19.6$~T. Since the $^{31}$P NMR measurements were carried out in a field of 9.394~T, the spin gap will be partially closed. The amount of gap that is expected to be closed at $H = 9.394$~T can be calculated as $\Delta^\prime/k_{\rm B} = \frac{H g \mu_{\rm B}}{k_{\rm B}} \simeq 12.5$~K. Thus, from $^{31}$P NMR at 9.394~T one should observe a spin gap of $\Delta/k_{\rm B}-\Delta^\prime/k_{\rm B} \simeq 13.6$~K, assuming a linear field dependence of $\Delta/k_{\rm B}$. Similarly, the corresponding values of the critical field for gap-closing and the gap at 9.394~T, in the case of a gapped Heisenberg chain are 15~T and 7.5~K, respectively.} Note, however, that the equation used to fit $K(T)$ is obtained in the low-field limit while our data are in a 9.394~T field which is comparable to the thermal energy below 10 K.


\subsubsection{\textbf{Spin-lattice relaxation rate, $1/T_1$}}
Spin-lattice relaxation rate $1/T_1$ provides direct access to the low-energy spin excitations, as it probes the nearly zero-energy limit (in the momentum space) of the local spin-spin correlation function.\cite{Moriya23} Therefore, spin-lattice relaxation time, $T_1$ is an important parameter which allows one to monitor the dynamics of a spin system in great detail. Since there are two inequivalent $^{31}$P sites in VP-$bpy$, it would be interesting to probe the spin dynamics by measuring $1/T_{1}$ separately at the two $^{31}$P sites. However, the peaks overlap in the spectrum over a large temperature range making independent determination of $1/T_1$ rather difficult.
\begin{figure}[h]
\includegraphics{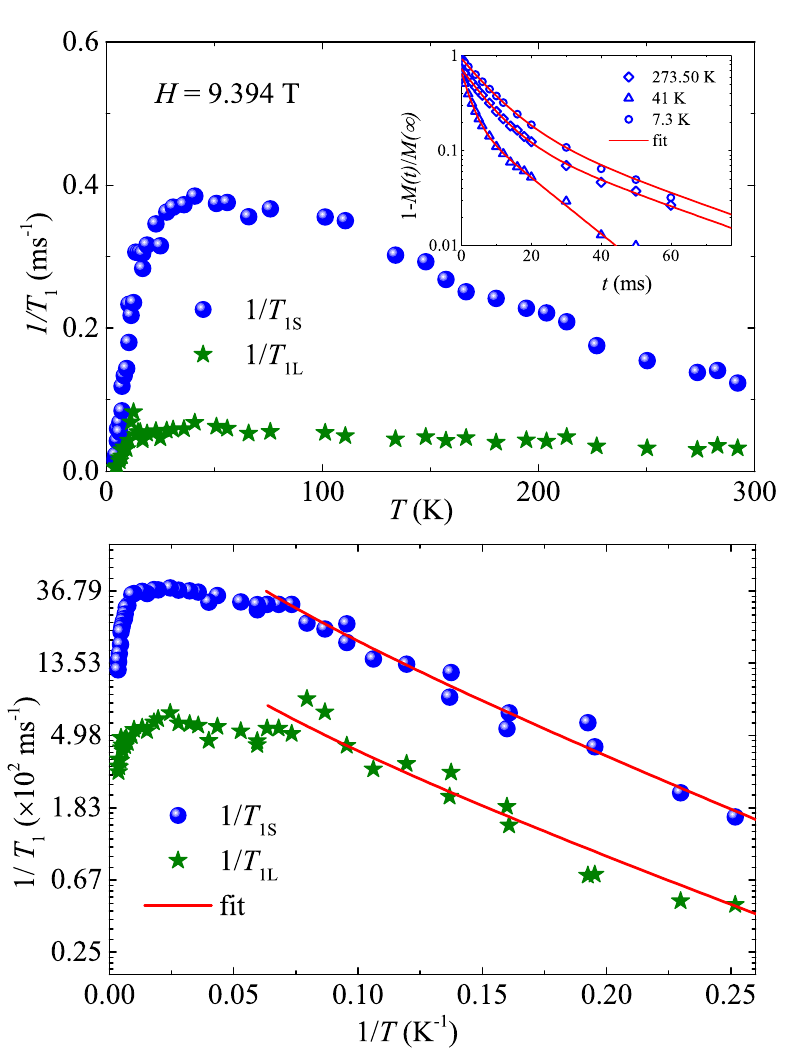}\\
\caption{Upper panel: $1/T_1$ ($1/T_{\rm 1S}$ and $1/T_{\rm 1L})$ is plotted as a function of temperature. Inset: Recovery of longitudinal magnetization as a function of waiting time $t$ at three different temperatures. Solid lines are the fits using Eq.~\eqref{Double_exp}. Lower panel: $1/T_{\rm 1}$ is plotted against $1/T$. The solid lines are the fits in the low-temperature region as discussed in the text.}
\label{Fig7}
\end{figure}
Therefore, our $T_1$ measurements were done at a frequency corresponding to the center of the spectra and using relatively narrow pulses. Recovery of the longitudinal magnetization at different temperatures after the saturation pulses was fitted using the double exponential function,
\begin{equation}\label{Double_exp}
1 - \frac{M(t)}{M(\infty)} = A_1 exp(-t/T_{\rm 1S})+ A_2exp(-t/T_{\rm 1L}).
\end{equation}
Here, $T_{\rm 1S}$ and $T_{\rm 1L}$ represent the shorter and longer components of $T_1$, respectively, $A_1$ and $A_2$ are the corresponding weight factors, and $M(t)$ and $M(\infty)$ are the nuclear magnetizations at time $t$ and at equilibrium ($t \rightarrow \infty$), respectively. The inset of Fig.~\ref{Fig7} presents the recovery curves at three different temperatures.

The $1/T_{\rm 1S}$ and $1/T_{\rm 1L}$ values estimated from the double exponential fit are plotted in Fig.~\ref{Fig7} as a function of temperature. Since there are two inequivalent P sites in the crystal structure, we suggest that $T_{\rm 1L}$ corresponds to the P(2) site which is weakly coupled while $T_{\rm 1S}$ corresponds to the P(1) site which is strongly coupled to the magnetic V$^{4+}$ spins.
At high temperatures, $1/T_{\rm 1L}$ is almost temperature-independent while $1/T_{\rm 1S}$ shows a gradual increase as the temperature is lowered. Below about 15~K, both $1/T_{\rm 1S}$ and $1/T_{\rm 1L}$ decrease rapidly towards zero because of the spin gap in the excitation spectrum.

Generally, $\frac{1}{T_1T}$ is expressed in terms of dynamical susceptibility $\chi(\vec{q},\omega)$ as\cite{Nath214430,Ranjith024422}
\begin{equation}
\frac{1}{T_{1}T} = \frac{2\gamma_{N}^{2}k_{B}}{N_{\rm A}^{2}}
\sum\limits_{\vec{q}}\mid A(\vec{q})\mid
^{2}\frac{\chi^{''}(\vec{q},\omega_{0})}{\omega_{0}},
\label{t1form}
\end{equation}
where the sum is over the wave vectors $\vec{q}$ within the first Brillouin zone,
$A(\vec{q})$ is the form factor of the hyperfine interactions as a
function of $\vec{q}$, and
$\chi^{''}_{M}(\vec{q},\omega _{0})$ is the imaginary part of the
dynamic susceptibility at the nuclear Larmor frequency $\omega _{0}$.
For $q=0$ and $\omega_{0}=0$, the real component $\chi_{M}^{'}(\vec{q},\omega _{0})$ corresponds to the uniform static susceptibility $\chi$. Usually, at high temperatures (i.e. in the paramagnetic regime) where spins are uncorrelated, $1/T_1$ is dominated by the uniform $q=0$ fluctuations and $1/T_1$ remains temperature independent.\cite{Moriya23} In such a scenario, one would expect $1/\chi T_1 T \sim$~constant. Surprisingly, in our compound $1/T_{\rm 1S}$ shows a strong temperature dependency and $1/\chi T_{\rm 1S}T$ is not constant, suggesting that the magnetic correlations ($q \neq 0$) persist upto high temperatures. This type of behaviour is typically observed in frustrated magnets. But $1/T_{\rm 1L}$ is almost temperature independent at high temperatures and $1/\chi T_{\rm 1L}T \sim$~constant. As we have discussed earlier, P(1) is located asymmetrically between two V(2) sites and is strongly coupled to the local moment fluctuations while P(2) is located symmetrically between two V(2) sites and is weakly coupled. Because of the symmetric location and weak hyperfine coupling, fluctuations from the neighbouring V(2) spins get nearly filtered out at the P(2) site which results in a nearly temperature independent $1/T_{\rm 1L}$ behaviour at high temperatures. On the other hand, due to asymmetric location and strong hyperfine coupling, the P(1) site experiences the persistent correlations at high temperatures giving rise to a temperature dependent $1/T_{\rm 1S}$ behaviour.

In order to estimate the spin gap, both $1/T_{\rm 1S}$ and $1/T_{\rm 1L}$ data in the low temperature region (below $\sim 15$~K) were fitted by the exponential function,\cite{Kageyama3168,Kikuchi6731}
\begin{equation}\label{T1_exp}
1/T_{\rm 1} \propto e^{-\Delta/k_{\rm B}T}.
\end{equation}
In the lower panel of Fig.~\ref{Fig7}, we plotted $1/T_{\rm 1S}$ and $1/T_{\rm 1L}$ vs $1/T$ where the $y$-axis is shown in log scale so as to highlight the activated behaviour at low temperatures. From the exponential fit, the value of spin gap is estimated to be $\Delta/k_{\rm B} \simeq 15.8$~K and 14.6~K from $1/T_{\rm 1S}(T)$ and $1/T_{\rm 1L}(T)$, respectively. These are consistent with the expected decrease due to the field, taking the low-field value from the $\chi(T)$ analysis with the dimer model.

Once again, for a gapped Heigenberg 1D system, $1/T_{\rm 1} \propto \Delta \sqrt{T} e^{-3\Delta/2k_{\rm B}T}$.\cite{Sachdev943,Damle8307} Applying this to our $1/T_{\rm 1L}$ data yields $\Delta/k_{\rm B} \simeq 8.3$~K while for the $1/T_{\rm 1S}$ data we got $\Delta/k_{\rm B} \simeq 7.5$~K (see the lower panel of Fig.~\ref{Fig7}). These are somewhat lower than those obtained from the $K(T)$ analysis but close to that expected from a linear decrease with applied field.\footnotemark[\value{footnote}] The static and dynamic properties observed in VP-$bpy$ are almost identical to the well-known spin dimer compound VO(HPO$_4$)0.5H$_2$O.\cite{Furukawa2393} NMR experiments are reported to show similar results as well on other gapped quantum spin systems such as Y$_2$BaNiO$_5$, CaV$_2$O$_5$, SrCu$_2$O$_3$, AgVP$_2$S$_6$, BaCu$_2$V$_2$O$_8$, (VO)$_2$P$_2$O$_7$, SrCu$_2$(BO$_3$)$_2$ etc.\cite{Shimizu9835,Iwase2397,Azuma3463,Takigawa2173,Ghoshray214401,Kikuchi6731,Kageyama3168}

\section{\textbf{Conclusion}}
We have studied the crystal structure and detailed magnetic properties of the quantum magnet VP-$bpy$. It shows neither a structural transition nor a change in symmetry down to 22.5~K. Our experimental results basically confirm VP-$bpy$ as a new spin-$1/2$ weakly coupled dimer compound. The nearest-neighbour V$^{4+}$ ions are coupled antiferromagnetically to form dimers with a singlet ground state. The analysis of $\chi(T)$ data with the dimer model establishes a spin gap of $\Delta/k_{\rm B} \simeq 26.1$~K between the singlet ground state and the triplet excited state at zero applied field and with a relatively weak inter-dimer coupling of $\sim 4.6$~K. The critical field corresponding to the gap closing is calculated to be about 19.6~T. On the other hand, fit of the susceptibility to the gapped 1D chain model gives $\Delta/k_{\rm B} \simeq 20$~K. The critical field required to close the gap in this case works out to about 15~T. Temperature dependent ESR intensity follows the same behaviour as $\chi(T)$. $^{31}$P NMR reveals two crystallographically inequivalent P sites in the compound. The strong hyperfine coupling [$A_{\rm {hf}} (1) \simeq 2963$~Oe/$\mu_{\rm B}$] for the P(1) site and weak hyperfine coupling [$A_{\rm {hf}} (2) \simeq 1466$~Oe/$\mu_{\rm B}$] for the P(2) site are consistent with the crystal structure.
The low temperature activated behaviour of NMR shift and $1/T_{1}$ unambiguously demonstrates a singlet ground state in the compound.
Our estimated values of $\Delta/k_{\rm B}$ from the low-field $\chi(T)$ data and $1/T_1$ analysis are close to what is expected at $H = 9.394$~T, assuming a linear decrease with field from its zero-field value.
These estimated parameters and the gapped nature make VP-$bpy$ a possible model compound for high field studies especially for exploring field-induced effects.

\section{\textbf{Acknowledgements}}
We would like to acknowledge IISER Thiruvananthapuram and IIT Bombay for providing the necessary experimental facilities.

%

\end{document}